\documentclass{PoS}
\usepackage{graphics}
\def \beq{\begin{equation}}
\def \eeq{\end{equation}}
\def \beqa{\begin{eqnarray}}
\def \eeqa{\end{eqnarray}}
\def \O{{\cal O}}
\def \ie{\emph{ i.e.\/}}
\def \etal{\emph{ et al.\/}}

\title{The critical end point of quantum chromodynamics}
\author{Rajiv Gavai\\
  Department of Theoretical Physics, Tata Institute of Fundamental
  Research,\\ Homi Bhabha Road, Mumbai 400005, India.\\
  E-mail: \email{gavai@tifr.res.in}}
\author{\speaker{Sourendu Gupta}\\
  Department of Theoretical Physics, Tata Institute of Fundamental
  Research,\\ Homi Bhabha Road, Mumbai 400005, India.\\
  E-mail: \email{sgupta@theory.tifr.res.in}}
\abstract{We investigate the critical end point (CEP) of QCD with
  two flavours of light dynamical quarks at finite lattice cutoff
  $a=1/4T$ using a Taylor expansion of the baryon number
  susceptibility. We find a strong volume dependence of the position
  of the critical end point. In the large volume limit we obtain
  $T^E/T_c\approx0.95$ and $\mu_B^E/T^E\approx1.1$, where $T_c$ is
  the cross over temperature at zero chemical potential, and $T^E$
  and $\mu_B^E$ are the temperature and the baryon chemical potential
  at the critical end point. The small value of $\mu_B^E$ places it
  in the range of observability in energy scans at the RHIC.}
\FullConference{XXIIIrd International Symposium on Lattice Field Theory\\
                25-30 July 2005\\
                Trinity College, Dublin, Ireland}
\ShortTitle{CEP of QCD}
\PoS{PoS(LAT2005)160}

\begin{document}
\section{Finding the CEP by Taylor Expansion}
In QCD with two flavours of quarks, the pressure and its Taylor expansion
are
\beq
   P(T,\mu_u,\mu_d) \equiv \left(\frac TV\right) \log Z(T,\mu_u,\mu_d)
                    = P(T,0,0) + \sum_{n_u,n_d} \chi_{n_u,n_d}
       \frac{\mu_u^{n_u}}{n_u!} \frac{\mu_d^{n_d}}{n_d!}
\label{press}\eeq
where $T$ is the temperature, $\mu_{u,d}$, the chemical potentials
for each flavour (since weak interactions are neglected, flavours
are exactly conserved), $V$, the volume, $Z$, the partition function,
and the Taylor expansion is around $\mu_u=\mu_d=0$. The Taylor
coefficients are called quark number susceptibilities (QNS). The
coefficients of order higher than 2 were called non-linear
susceptibilities (NLS) in \cite{defnls}.  Most lattice computations
are performed in the flavour symmetric limit, $m_u=m_d$, which
forces the coefficients to have the symmetry $\chi_{nm}=\chi_{mn}$.
Also, CP symmetry forces $\chi_{n,N-n}=0$ whenever $N$ is odd.

We are interested in the critical end point in the plane of $T$ and
the baryon chemical potential $\mu_B=3\mu_u=3\mu_d$. The second
derivative
\beq
   \chi_{20}(T,\mu_B) = \left.\frac{\partial^2 P}{\partial\mu_u^2}
        \right|_{\mu_u=\mu_d=\mu_B/3}
\label{qns}\eeq
diverges at the critical point in the thermodynamic limit. We
construct the Taylor series for this QNS from (\ref{press}).
In \cite{cep} we extract the Taylor coefficients of $\chi_{20}(T,\mu_B)$
from lattice simulations in order to estimate the CEP. Note that
we scan along lines of constant $T$. At the CEP the Taylor expansion
in (\ref{qns}) would break down. Hence, the radius of convergence
of the series, in the thermodynamic limit, would give the position
of the CEP provided that there is no other critical point nearer
to $\mu_u=\mu_d=0$.

Our analysis involves examination of finite volume effects in
$\chi_{20}(T,\mu_B)$. On any finite volume, as one scans along a line
of constant $T$, one sees a peak in the QNS. With increasing volume
one sees larger and sharper peaks, which go smoothly into the
non-analytic divergence in the thermodynamic limit. This implies
an interesting behaviour for the Taylor coefficients on finite
volumes. Estimates of the radius of convergence from terms up to
some order $N_;(V)$ should indicate a finite value; but terms of
order greater than $N_;(V)$ would show a growth in the radius of
convergence. With increasing $V$ one should observe $N_;(V)$ to
be increasing. Estimates of the radius of convergence have to be
performed, perforce, at finite $V$, and therefore only the
estimates for orders less than $N_;(V)$ should be used, and continued
to the thermodynamic limit using standard finite size scaling methods.

The main systematic errors in the location of CEP by lattice methods
are due to three sources--- lattice spacing effects, finite volume
effects, quark mass effects. In this work we show that finite volume
effects can be controlled when ${\cal V}=m_\pi V^{1/3}>5$. By working in the
large volume region and taking small quark masses such that
$m_\pi/m_\rho=0.3$ (close to the physical value) we are able to
control the last two sources of errors simultaneously for the first
time. Lattice spacing errors will be considered later. Our simulation
parameters are listed in \cite{cep}. We expand $P$ to 8th order in
$\mu_B$.

\section{Optimizing the computations}

\begin{figure}
\begin{center}
   \scalebox{0.4}{\includegraphics{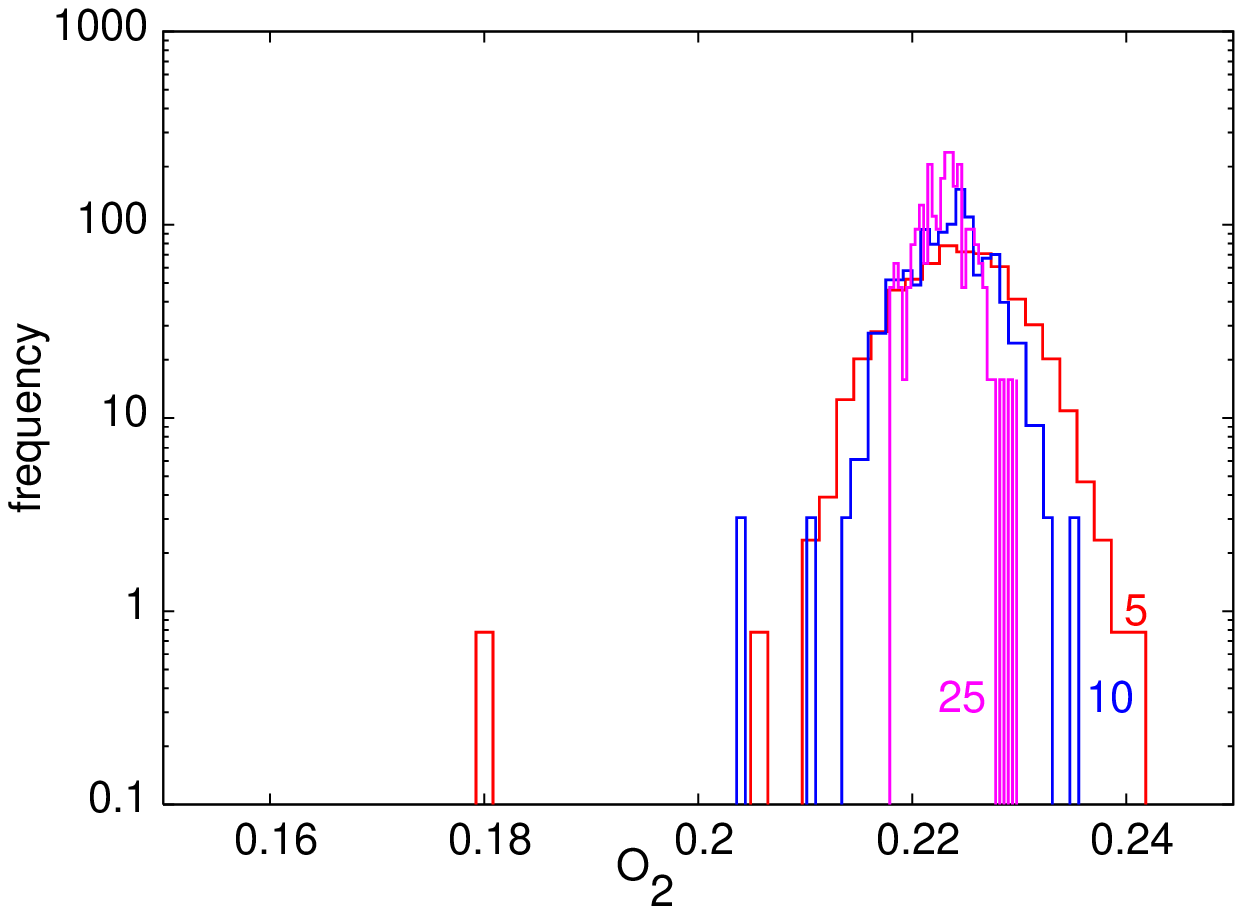}}
   \scalebox{0.4}{\includegraphics{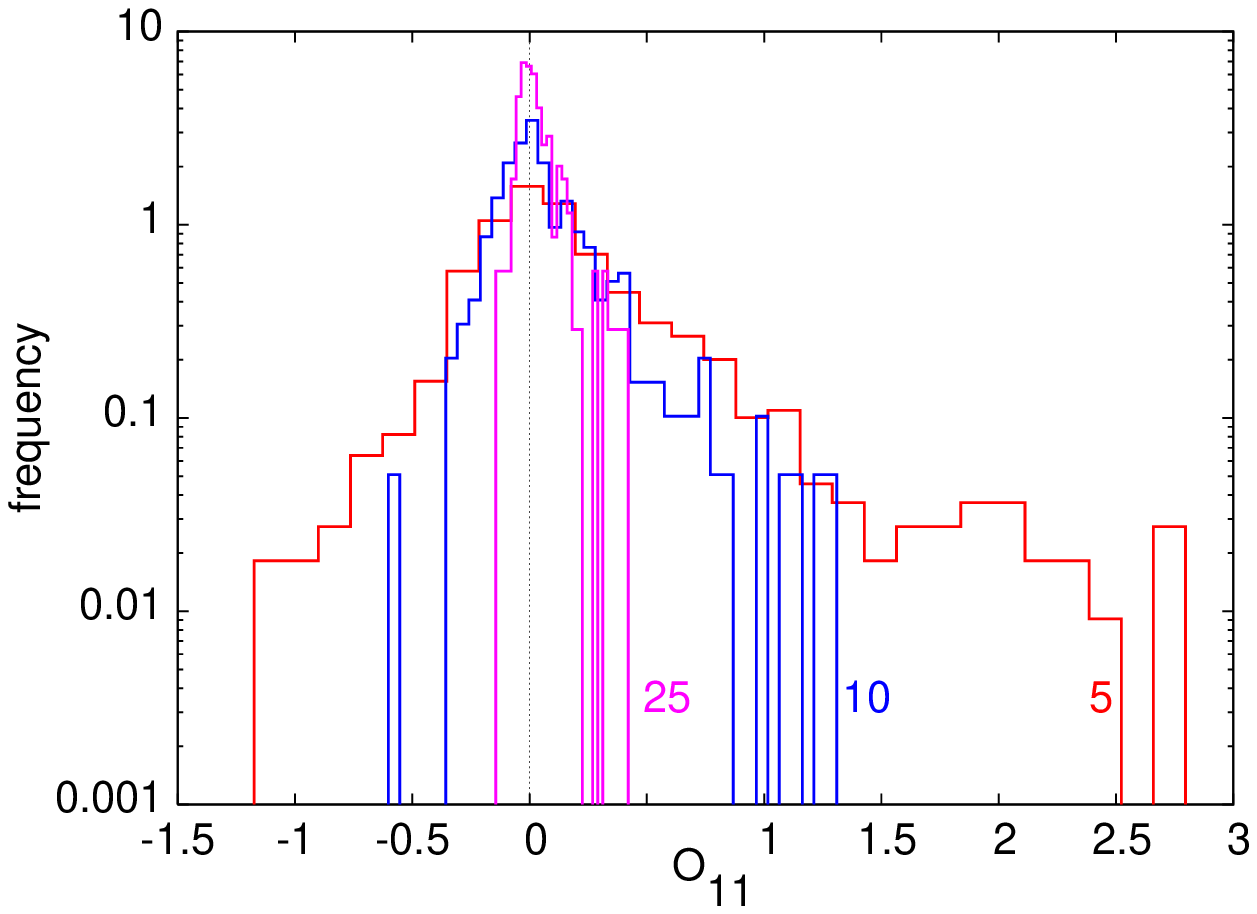}}
\end{center}
\caption{The distribution of estimators of the traces involved in
  (a) $\chi_{20}$ and (b) $\chi_{11}$ on one configuration at $0.8T_c$ by
  Gaussian distributed noise vectors. The different histograms in
  each case are labeled by the number of vectors used to make one
  measurement. Noise in the measurement of $\chi_{11}$ arises from
  the significantly larger spread of the second (highly non-Gaussian)
  histogram with a much smaller mean, and therefore requires many more
  vectors to control.}
\label{fg.distrib}\end{figure}

Every derivative of $\log Z$ with respect to $\mu$ lands on the
Dirac determinant, since this is the part of the measure which
contains $\mu$.  Every derivative of the determinant of a matrix
creates an inverse power of the matrix, \ie, a quark propagator in
this example. Thus, when Taylor series are examined at order $N$,
they may contain terms containing up to $N$ propagators. On every
gauge configuration one is therefore required to construct fermion
traces with many propagators, implying a need for multiple Dirac
matrix inversions. The way to optimize the number of inversions
is to map this problem on to a computer science problem
called the ``Steiner problem'' \cite{charikar}. Details are given
in \cite{cep}.

Fermion traces are obtained using the usual stochastic method---
random vectors are drawn from some ensemble (usually Gaussian or
$Z_2$), the expectation of the operator is found on each vector
and the average is found over the noise ensemble. Optimization
involves choosing the ensemble and the number of vectors. We found
that for this specific purpose the Gaussian ensemble is superior
to the $Z_2$ noise ensemble. The number of vectors needed increases
with the order, and at each order Fermion-line-disconnected operators
require more vectors than connected operators.

Since multiple matrix inversions are required to on each gauge
configurations, one may expect a priori that some preconditioning
or reuse of vectors may improve the performance of the inverter.
However, it turns out that the extra arithmetic effort involved in
this is comparatively large, and leads to gains of around 10\% (with
a less sparse Dirac operator, \ie, with improved actions, there may
be more gain). We therefore did not optimize the inversion in this
manner. The main optimization of the CG inverter was to tune the
stopping criterion.

We measured the autocorrelations of the Wilson line and the quark
condensate during the R-algorithm run, and chose to analyze only
one configuration per autocorrelation time, gathering statistics
of 50--100 configuration at each coupling on every lattice size.
Since autocorrelations are large in the vicinity of the QCD crossover
temperature, $T_c$, it took massive computational effort to generate
sufficient statistics in this interesting region. The effort required
in the measurements was significantly smaller.

\section{Non-linear susceptibilities}

\begin{figure}
\begin{center}
   \scalebox{0.5}{\includegraphics{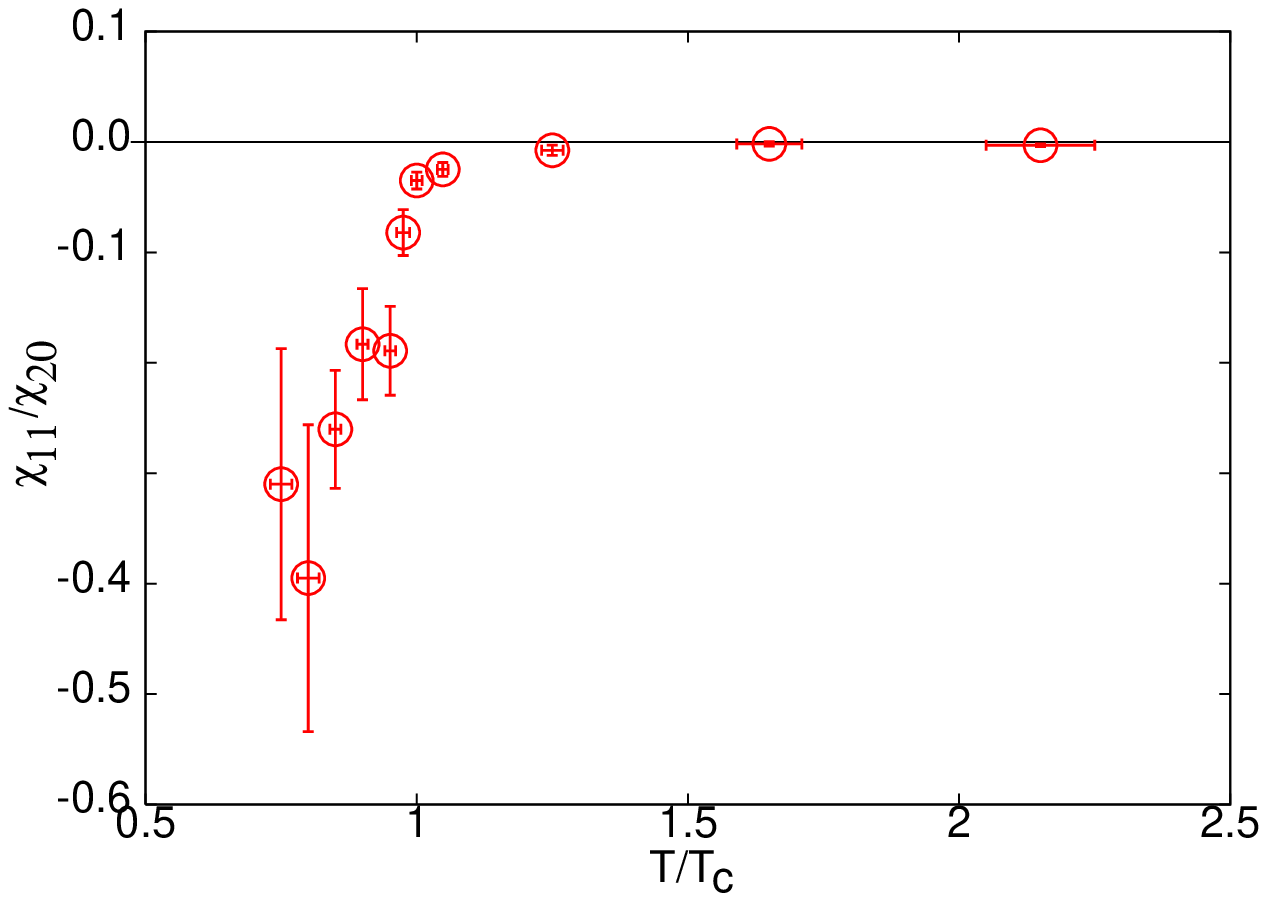}}
   \scalebox{0.5}{\includegraphics{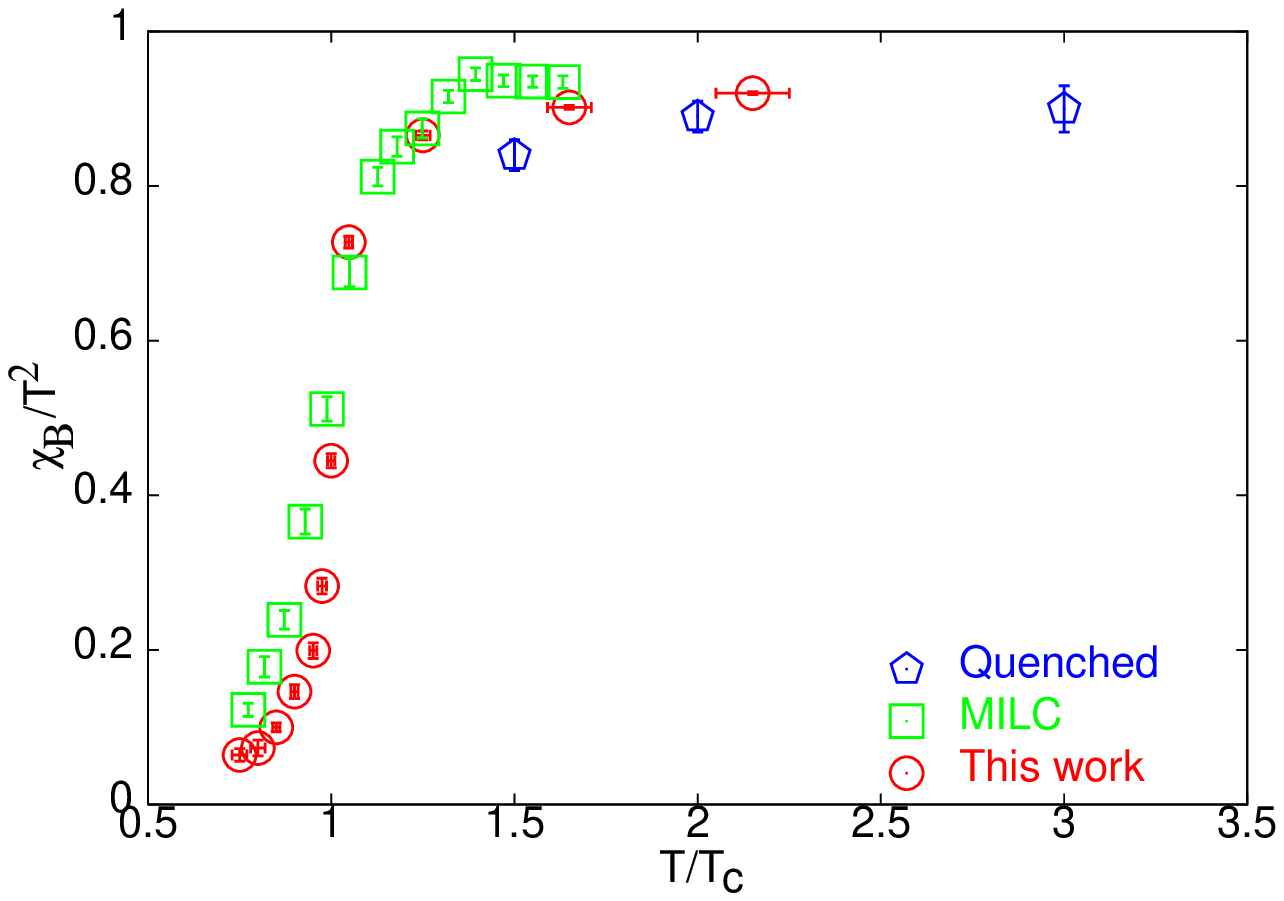}}
\end{center}
\caption{The first panel shows the ratio $\chi_{11}/\chi_{20}$ as
  a function of $T/T_c$. The second shows the scaled value of
  $\chi_{20}$ from our computation (see text for details) compared
  to data from \cite{milc} and quenched QCD \cite{valence}.}
\label{fg.sign}\end{figure}

Different systematics are seen for non-linear susceptibilities (NLS)
in the three regions, well below $T_c$, near $T_c$ and well above
$T_c$ \cite{nls}.  An example is provided by the ratio
$\chi_{11}/\chi_{20}$ which measures the importance of sign
fluctuations in the partition function at small chemical potential.
This is of order unity at $0.75T_c$, and decreases rapidly in the
region near $T_c$, becoming of the order of $10^{-3}$ at large $T$
(see Figure \ref{fg.sign}).  $\chi_{20}$ itself crosses over from
a small value below $T_c$ to a much larger values above $T_c$. In
Figure \ref{fg.sign} the value shown is the observed value scaled
by a factor which would transform measurements of $\chi_{20}$ in
the quenched theory at the same lattice spacing into the continuum.
We have argued before that this procedure is within 5---10\% of the
true continuum limit. Taking into consideration this uncertainty,
and differences in quark masses and lattice volumes, the values are
in agreement with the $N_t=8$ computations with improved staggered
quarks of \cite{milc}.

Well above $T_c$ there is a hierarchy of values of the NLS which
is consistent with weak-coupling power counting rules. In the
vicinity of $T_c$ these rules break down, and some of the NLS peak.
These peaks are found to be due to one particular class of operators,
which measure fluctuations of the fermion-line-connected operator,
$\O_2$, contributing to $\chi_{20}$ \cite{nls}.

\section{The CEP}

\begin{figure}
\begin{center}
   \scalebox{0.5}{\includegraphics{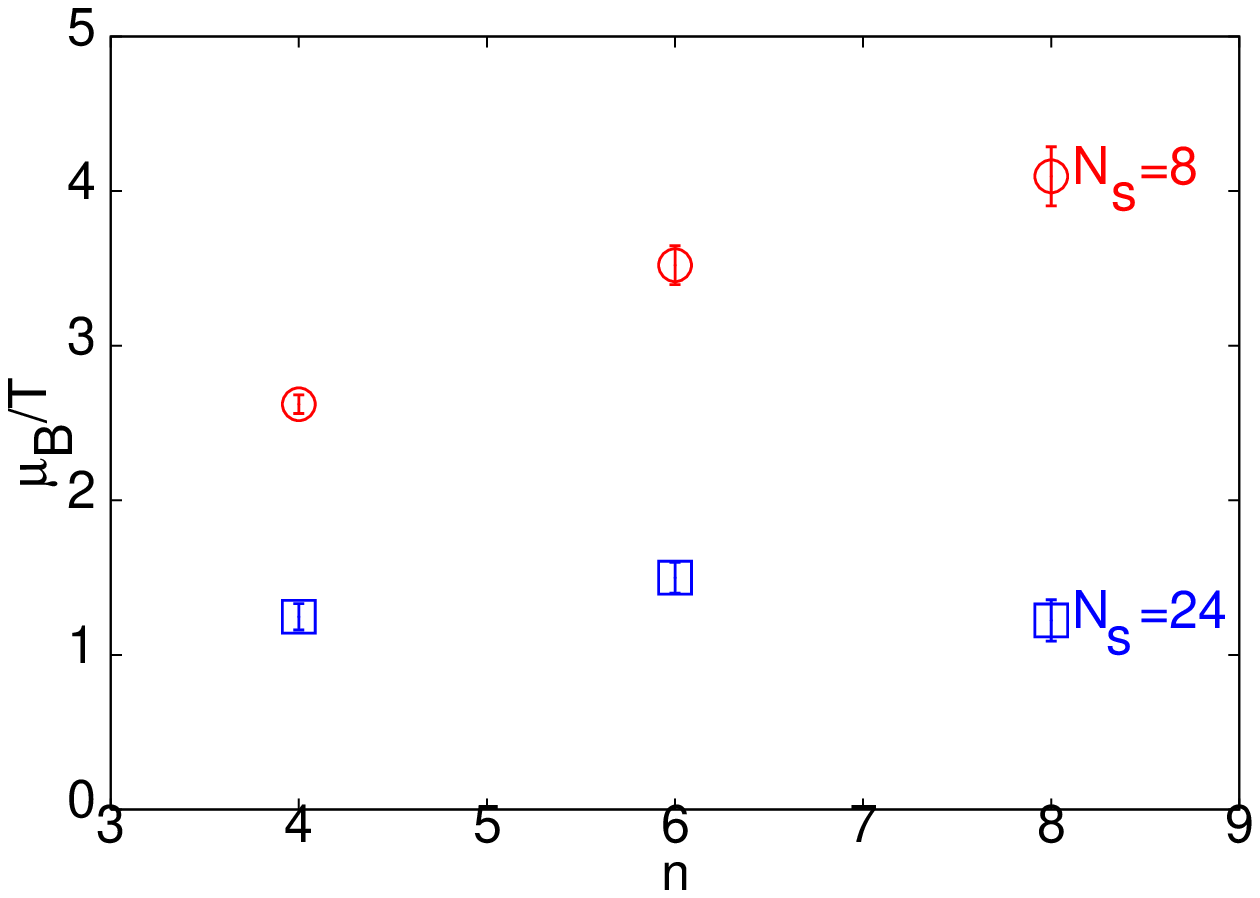}}
   \scalebox{0.5}{\includegraphics{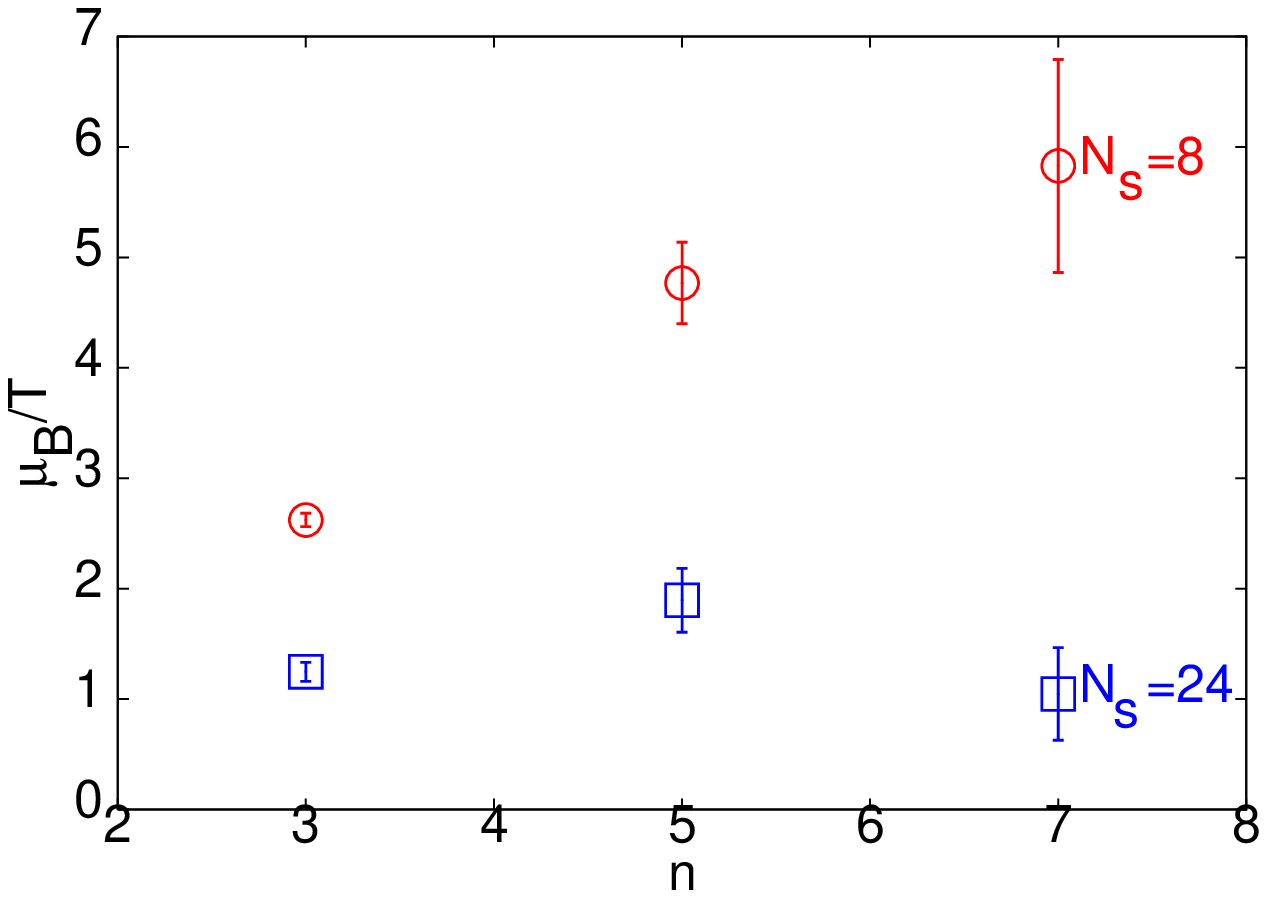}}
\end{center}
\caption{The two estimators of the radius of convergence of the series for
   $\chi_{20}(\mu_B)$--- (a) $\rho_{2n}$ and (b) $r_{2n}$. The nature of
   the finite volume effects are clear.}
\label{fg.radii}\end{figure}

\begin{figure}
\begin{center}
   \scalebox{0.5}{\includegraphics{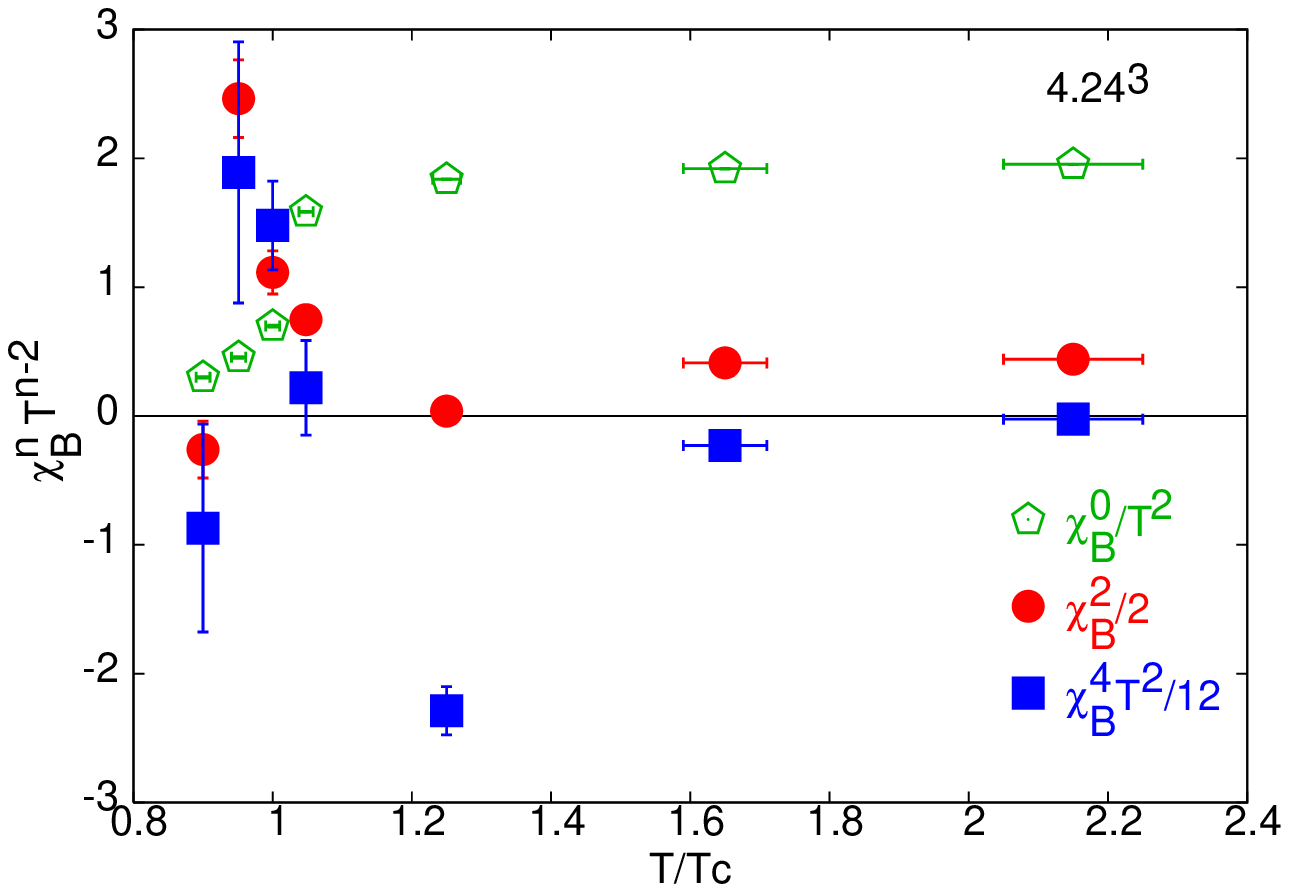}}
   \scalebox{0.5}{\includegraphics{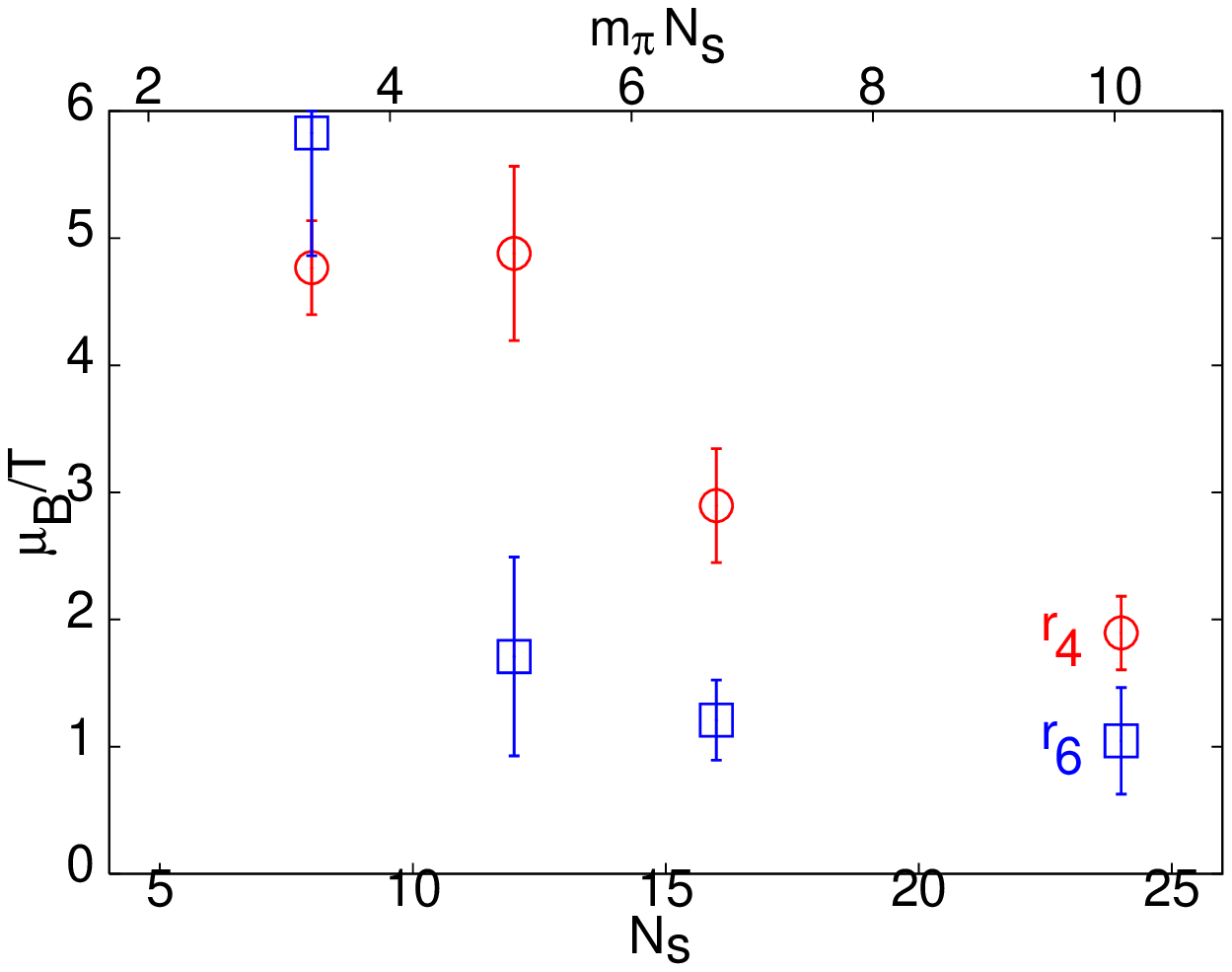}}
\end{center}
\caption{The first panel shows the Taylor coefficients of $\chi_{20}$ at
   orders 0, 2 and 4 as a function of $T/T_c$. The second shows $r_4$ and
   $r_6$ at $T=0.95t_c$ as a function of $V^{1/3}$ and $\cal V$.}
\label{fg.somemore}\end{figure}

\begin{figure}
\begin{center}
   \scalebox{0.5}{\includegraphics{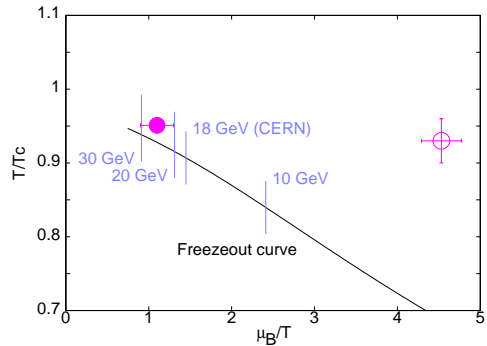}}
\end{center}
\caption{Our evaluation of the critical end point for QCD in the
   thermodynamic limit compared to an earlier evaluation \cite{fk}.
   Also shown is the freezeout curve in heavy-ion collisions with
   the center of mass energy marked. It seems that energy scans at
   RHIC may see interesting physics.}
\label{fg.cep}\end{figure}

Given the Taylor expansion of an even function $f(x)=\sum_n f_{2n}x^{2n}$ 
around the symmetric point $x=0$, one may define the radius of convergence,
$x_*$, in several equivalent ways. Two definitions which we use here are
the limits of the successive approximants
\beq
   \rho_{2n}=\left|\frac{f_0}{f_{2n}}\right|^{1/2n} \quad{\rm or}\quad
   r_{2n}=\sqrt{\left|\frac{f_{2n}}{f_{2n+2}}\right|}
\label{rads}\eeq
The results for $\chi_{20}(\mu_B)$ evaluated on two different volumes are
shown in Figure \ref{fg.radii} at $T=0.95T_c$. Note that the volume effects
are compatible with the expectations discussed in Section 1.

One important question is whether the series diverges for real
$\mu_B$ or imaginary $\mu_B$. If it is the first possibility which
is realized, then all terms in the series should be positive; if
the second, then the series should alternate. We find that all the
terms are positive in the range $0.8\le T/T_c\le1$, within which
the CEP is located (see the first panel of Figure \ref{fg.somemore}),
indicating that the first possibility is actually realized. For
$N_f=2$ the only other nearby critical point is that at
$\mu_I/T_c=m_\pi/T_c=1.6$ for $T=0$. This is further away from the
origin than the critical point identified by the radius of convergence.
Hence we conclude that the radius of convergence identifies the
CEP.

Finite size effects are of two kinds. At small volumes, \ie,
when ${\cal V}<5$ (in this formula the appropriate
value of $m_\pi$ to use is that measured in very large volumes at $T=0$),
the finite volume effects are dominated by the infrared cutoff on
the Dirac eigenvalues presented by the volume. Thermodynamics is
recovered when ${\cal V}>5$. We show the effects of this crossover
in the second panel of Figure \ref{fg.somemore}. In the large volume
portion one may use standard (thermodynamic) finite size scaling
analysis (using the expected Ising exponents) to continue the results
to infinite volume.

Our final result on the CEP are displayed in Figure \ref{fg.cep}.
Also shown for comparison is an older evaluation of this point by
a different method in $N_f=2+1$ QCD with the same pion mass and a
volume ${\cal V}\le3.9$ \cite{fk} (other approaches include
\cite{maria,owe,karsch,azcoiti}).
Since our results at the comparable volumes agree with this evaluation,
we believe that the discrepancy cannot be attributed to the difference
in the flavour content of the sea.  Also shown for comparison is
the freezeout curve in heavy-ion collisions. It seems that the CEP
may be observed in energy scans at the RHIC.

 \end{document}